\documentclass{llncs}

\usepackage[british]{babel}
\usepackage{amsfonts}
\usepackage{booktabs}
\usepackage{hyperref}
\usepackage{svn-multi}
\usepackage{url}
\usepackage{xspace}
\usepackage{graphicx}
\usepackage{tikz}
\usetikzlibrary{calc}
\usetikzlibrary{positioning}
\usetikzlibrary{automata}
\usetikzlibrary{shapes}
\usetikzlibrary{decorations.pathreplacing}
\usepackage{wrapfig}
\usepackage{times}
\usepackage{amsmath}

\usepackage[caption=false]{subfig}
\usepackage{listings}
\newcommand{\doi}[1]{\href{#1}{\nolinkurl{https://doi.org/#1}}}

\title{How Testable is Business Software?}

\author{
  Peter Schrammel\inst{1,2}\orcidID{0000-0002-5713-1381}
}

\institute{
  ~\inst{1}Diffblue Ltd, Oxford, UK \hspace*{3mm}
  ~\inst{2}University of Sussex, Brighton, UK
}

\begin{document}

\maketitle

\begin{abstract}
Most businesses rely on a significant stack of software to perform
their daily operations. This software is business-critical as defects
in this software have major impacts on revenue and customer
satisfaction.  The primary means for verification of this software is
testing.  We conducted an extensive analysis of Java software packages
to evaluate their unit-testability.  The results show that code in
software repositories is typically split into portions of very trivial
code, non-trivial code that is unit-testable, and code that cannot be
unit-tested easily.  This brings up interesting considerations
regarding the use of test coverage metrics and design for testability,
which is crucial for testing efficiency and effectiveness. Lack of
unit-testability is an obstacle to applying tools that perform
automated verification and test generation. These tools cannot make up
for poor testability of the code and have a hard time in succeeding or
are not even applicable without first improving the design of the
software system.
\end{abstract}

\section{Introduction} \label{sec:introduction}

Enterprise software is business-critical as any bugs in
that software have a tremendous impact on revenue, customer
satisfaction, reputation, competitiveness, etc.
Companies are under constant pressure to deliver features
ever faster and cheaper. And defects should be detected early in the
development process to keep the cost of fixing them low.  The cost of
a bug that causes damage to a customer is orders of magnitude higher
than a bug that a developer notices before they commit the code into
the code base.  This reasoning is not much different from what we are
used to in the safety-critical domains.

\begin{figure}[t]
\centering  
\includegraphics[width=0.25\textwidth]{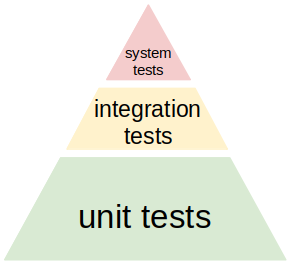}  
\caption{\label{fig:test-pyramid}
Testing pyramid (cf.~\cite{Coh09})
}
\end{figure}

Testing plays the key role in
detecting defects and ensuring high quality levels.  However,
different sorts of tests play different roles.  In order to
‘shift-left’ the defect detection, tests are needed that can run early
in the development process.  These tests must be fast so that
developers can run them on their machines and on every pull request.
Having a solid base of unit tests (as shown in
Figure~\ref{fig:test-pyramid}) is best-suited to deliver on this.
Integration and system tests are essential, but if there are too many
of them then they cannot be run continuously on each pull request any
more.

This paper reports on experiences made in developing commercial tools
for automated test generation of Java applications.
Usually, scalability and state space explosion are mentioned
as primary problems encountered in applying automated verification
techniques to real-world code bases.
However, also testability is a huge issue and affects the performance
or even applicability of these tools.
We encountered cases where none of our improvements to our automated
test generation tool showed the expected results.
We then manually analyzed the methods in the code base for which automated
test generation did not produce results.
These analyses exhibited that it is often difficult -- even for a human --
to write unit tests for certain parts of the code. Spending two hours and
more on trying to write a single test and eventually failing to do so
is not exceptional and a sign for severe testability issues.
Often the setup of tests becomes horrendously complex and dependencies
cannot be mocked appropriately.

\begin{figure}[t]
\centering  
\includegraphics[width=0.8\textwidth]{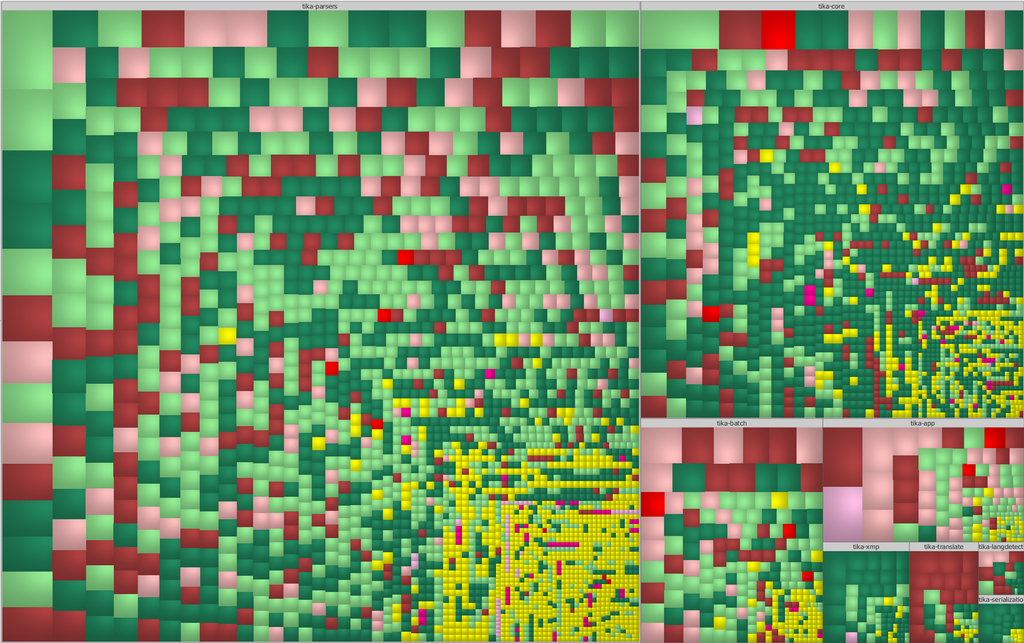}  
\caption{\label{fig:tika-testability-map}
Unit testability map for a code base
}
\end{figure}

Figure~\ref{fig:tika-testability-map} shows a \emph{unit testability map}
of a code base.%
\footnote{\url{https://github.com/apache/tika}}
Each box is a method. Its size is proportional to the lines of code.
Green colors indicate that the method is unit-testable.  Red colors
indicate that the method has some testability challenges.  And yellow
colors indicate that the method is very trivial, like a getter or
setter, one would not usually write a test for. The larger groups of blocks
are the Maven modules of the project.%
\footnote{The different shades of colors indicate different reasons
for testability issues, which we will explain later.}
Looking at the map it becomes apparent that there are some methods that
are difficult to test; and some modules have more problematic methods
than others.
In this paper we will report on how to produce
such a detailed assessment of unit testability together with diagnostic
information for individual methods, with the ultimate goal
to improve the unit testability of a code base and increase the
performance of automated test generation tools.

We will also discuss some learnings that we draw from this experience
since testability deficiencies may also impact other automated
verification techniques, not just test generation.
Business software is a high-value assets for any company and these
code bases are constantly being improved and adapted to business needs.
Design for testability as a concept has been known for
decades~\cite{Bin94,Bin99}, but it does not seem to be fully standard
practice yet.
Any automation support that can be provided to support improved
testability may deliver high impact.

\paragraph{Outline}
After explaining some preliminaries, we will present
\begin{itemize}
\item an illustration of unit testability challenges, in particular
  concerning mockability, commonly occurring in Java code bases
  (Section~\ref{sec:mockability}),
\item a unit testability model and corresponding static analysis that we implemented (Sections~\ref{sec:model} and~\ref{sec:analysis}),
\item results of an experimental evaluation on several millions lines of Java code (Section~\ref{sec:results}), and
\item a discussion of impacts of unit testability on testing
  efficiency, the role of design for testability and implications of
  testability limitations on applying automated verification tools to
  business software (Section~\ref{sec:implications}).
\end{itemize}

\section{Preliminaries} \label{sec:preliminaries}

Before we start, we clarify the concepts of unit test, critical code,
and testability.

\subsection{Unit Tests} \label{sec:unittesting}

A unit test tests a simple scenario, usually a single method call on a
small testable unit such as a class.
As shown in Figure~\ref{fig:unit-test}, unit tests usually have three
sections: first the inputs to the method under test (MUT) are
\emph{arrange}d. Then the MUT is called in the \emph{act} section.
Finally, the effects of the MUT execution are \emph{assert}ed.
Dependencies to other subsystems are usually mocked; external
dependencies such as socket connections are definitely mocked. For
example, a unit test usually does not talk to a database system.
Unit tests are expected to run very fast (in the order of a few
milliseconds).  They have no effects on other tests so that they can
be run in parallel without interfering.

\begin{figure}[t]
\begin{lstlisting}
public class ProductTest {
  @Test 
  public void testSend() {
    // Arrange the inputs and mocks
    Product product = new Product();
    product.addExpiryDate();
    // Act: call the method under test (MUT)
    boolean isExpired = product.isExpired();
    // Assert on the effects
    assertTrue(isExpired);
  }
}
\end{lstlisting}
\caption{\label{fig:unit-test}
Arrange-Act-Assert structure of a unit test
}
\end{figure}

The arrange-act-assert model is quite schematic, but
gives some key how to read these tests.
Real-world tests look more like the one in
Figure~\ref{fig:unit-test-alfresco}, but one can still identify the
three sections to some degree.

\begin{figure}[t]
\begin{lstlisting}
@Test
public void testPropertyMappingGlobalOverride()
    throws Exception {
  String propertyPrefix =
    AbstractMappingMetadataExtracter.PROPERTY_PREFIX_METADATA +
    DummyMappingMetadataExtracter.EXTRACTER_NAME +
    AbstractMappingMetadataExtracter
      .PROPERTY_COMPONENT_EXTRACT;

  ApplicationContext ctx =
    MiscContextTestSuite.getMinimalContext();
  Properties globalProperties =
    (Properties) ctx.getBean("global-properties");
  globalProperties.setProperty(
    propertyPrefix + "namespace.prefix.my",
    DummyMappingMetadataExtracter.NAMESPACE_MY);
  globalProperties.setProperty(
    propertyPrefix + DummyMappingMetadataExtracter.PROP_A,
    " my:a1, my:a2, my:c ");
  
  extracter.setApplicationContext(ctx);
  
  extracter.register();
  // Only mapped 'a'
  destination.clear();
  extracter.extract(reader, destination);

  assertEquals(
    DummyMappingMetadataExtracter.VALUE_A,
    destination.get(DummyMappingMetadataExtracter.QNAME_C));
}
\end{lstlisting}
\caption{\label{fig:unit-test-alfresco}
A typical real-world unit test (cf. \url{https://github.com/Alfresco/alfresco-repository})
}
\end{figure}

\subsection{Critical Code} \label{sec:critical}

\begin{figure}[t]
\centering  
\includegraphics[width=0.40\textwidth]{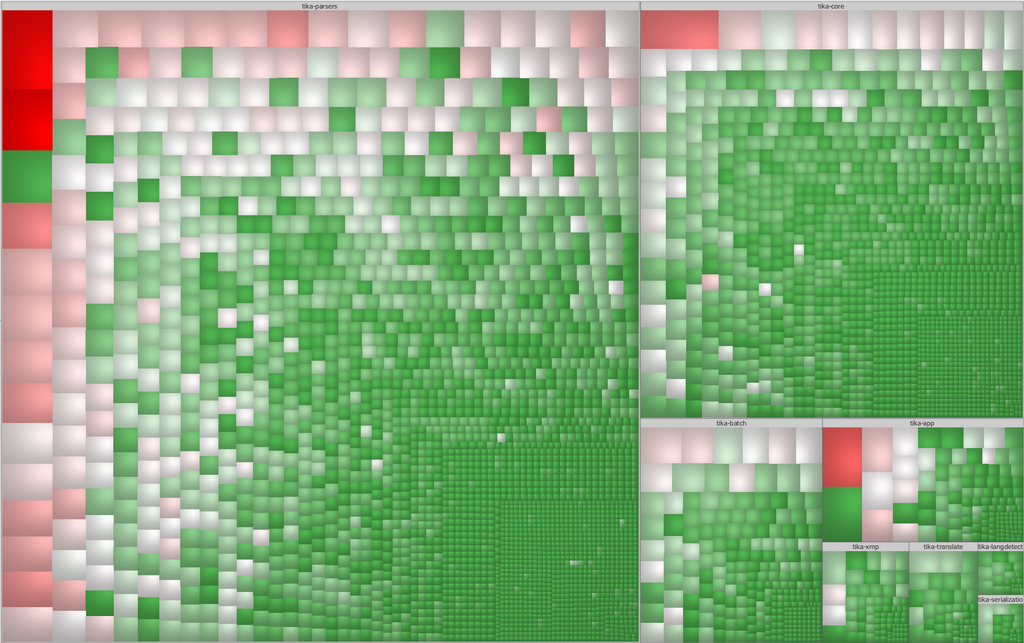}
\caption{\label{fig:tika-complexity-map}
Complexity map for a code base
}
\end{figure}

Not all code in a code base is equally critical.
When talking to software architects and engineering managers, they
are primarily concerned with having tests for the critical business
logic.
For a developer on a code base it is quite self-evident what is critical and
what is not.  For a tool this is less obvious.
Often cyclomatic complexity~\cite{McCabe1976ACM} is used as a proxy
for spotting where the hard parts are.
Figure~\ref{fig:tika-complexity-map} shows a complexity map of
a code base.%
\footnote{\url{https://github.com/apache/tika}}
The less green and more red, the more complex.
We will not use cyclomatic complexity, though, but a simple analysis
to identify code that is really trivial such as getters and
setters. One would not write explicit tests for such methods, but they
are necessary to write any other test because they are required to set
up objects and access the state when writing assertions.
Everything else might be critical and we will not make a
judgment on how critical it is.

Once we know the code that we consider critical we
will distinguish between unit-testable and not unit-testable code.  We
will perform an under-approximation - so methods might be classified
as unit-testable, although they have testability challenges.
We will discuss this analysis in Section~\ref{sec:analysis}).

\begin{figure}[t]
\centering  
\includegraphics[width=0.40\textwidth]{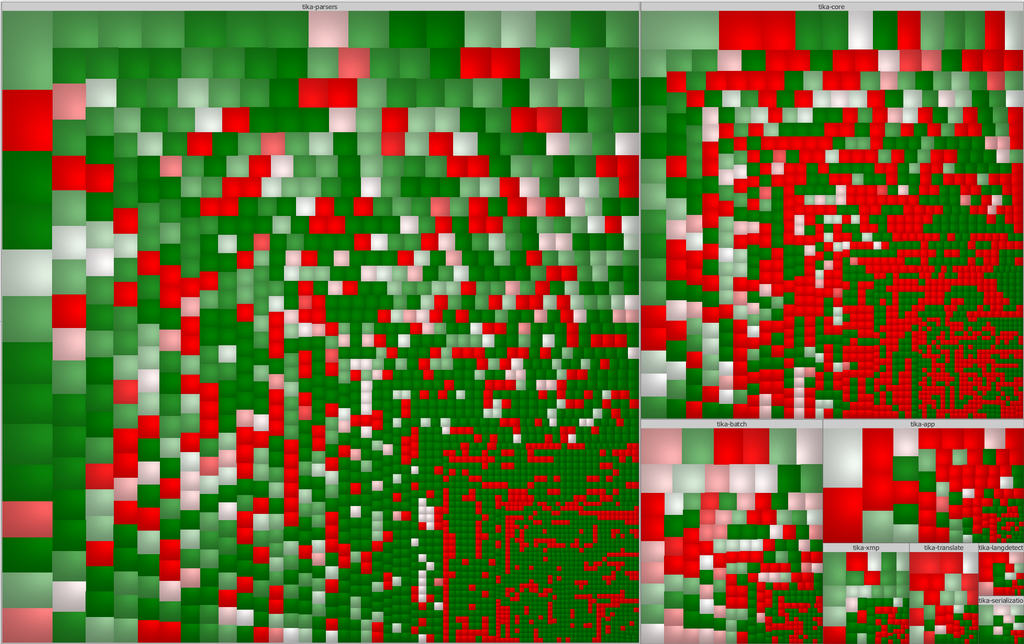}
\caption{\label{fig:tika-coverage-map}
Coverage map for a code base
}
\end{figure}

Beyond that, one could also look at the existing tests and evaluate
their composition unit vs integration vs system tests, measure their
coverage and adequacy, etc.
Figure~\ref{fig:tika-coverage-map} shows a coverage map of
a code base.
One can see that quite some effort was actually made to cover
particularly the more complex parts.
In the sequel, we will focus on how to distinguish between
unit-testable and not unit-testable code, though.

\begin{figure}[t]
\centering
\begin{tabular}{l@{\qquad}l}
\begin{lstlisting}
public class Product {
  private LocalDateTime expiryDate;

  public void addExpiryDate() {
   this.expiryDate = LocalDateTime.now()
     .plus(30, DAYS);
  }
  public boolean isExpired() {
    return this.expiryDate
      .isBefore(LocalDateTime.now());
  }
}
\end{lstlisting} &
\begin{lstlisting}
public class ProductTest {
  @Test public void testSend() {
    // Arrange
    Product product = new Product();
    product.addExpiryDate();
    Thread.sleep(31*24*3600); // ???
    // Act & Assert
    assertTrue(product.isExpired());
  }
}
\end{lstlisting}
\end{tabular}
\caption{\label{fig:mockability-controllability}
Controllability aspect of mockability
}
\end{figure}

\begin{figure}[t]
\centering
\begin{tabular}{l@{\qquad}l}
\begin{lstlisting}
public class Product {
  private LocalDateTime expiryDate;
  private Clock clock = Clock.systemUTC();
  public void addExpiryDate() {
    this.expiryDate = LocalDateTime
      .now(clock).plus(30, DAYS);
  }
  public boolean isExpired() {
    return this.expiryDate
      .isBefore(LocalDateTime.now(clock));
  }
  void setClock(Clock clock) {
    this.clock = clock;
  }
}
\end{lstlisting} &
\begin{lstlisting}
public class ProductTest {
  @Test public void testExpired() {
    // Arrange
    Product product = new Product();
    product.setClock(
      Clock.fixed(Instant.EPOCH));
    product.addExpiryDate();
    product.setClock(Clock.fixed(
      Instant.EPOCH.plus(31, DAYS)));
    // Act & Assert
    assertTrue(product.isExpired());
  }
}
\end{lstlisting}
\end{tabular}
\caption{\label{fig:mockability-controllability-solved}
Controllability aspect of mockability: dependency injection
}
\end{figure}

\subsection{Testability} \label{sec:testability}

Nate Edwards wrote 1975 on software testability:
\emph{``If modularity is controlled so that the function of a module is independent of the source of its input, the destination of its output, and the past history of use of the module, the difficulty of testing the modules and structures assembled from the modules is greatly reduced.''}~\cite{Edwards1975TheEO}
In other words, the more modular, functional, less coupled and less stateful the easier software is to test.

Roy Freedman made this more precise and defined it as the application of Kalman’s observability and controllability concepts~\cite{Klmn1959OnTG} to software.
\emph{``The concept of [...] testability of software is defined by applying the concepts of observability and controllability to software. It is shown that a [...] testable program does not exhibit any input-output inconsistencies and supports small test sets in which test outputs are easily understood. Metrics that can be used to assess the level of effort required in order to modify a program so that it is [...] testable [...].''}~\cite{Freedman1991TestabilityOS}

Robert Binder then put it into a nutshell:
\emph{``Testability has two key facets: controllability and observability. To test a component, you must be able to control its input (and internal state) and observe its output. If you cannot control the input, you cannot be sure what has caused a given output. If you cannot observe the output of a component under test, you cannot be sure how a given input has been processed.''}~\cite{Bin94}

The two main aspects of testability are controllability and observability.
\begin{itemize}
\item \emph{Controllability} in a control system means that we can steer it into any desired state by applying specific inputs.
For software, it means the ability to arrange the inputs of the method under test to make it take a desired code path.
This includes also the ability to control the effects of dependent components, which we will call mockability (see Section~\ref{sec:mockability}).
Controllability is affected by non-determinism, essentially anything that talks to the operating system such as network, file system, getting the system time, random generators, and multi-threading. Also unreachable code that cannot be exercised plays a role.
\item \emph{Observability} in a control system means to be able to determine the state from the outputs.
In the context of software testing it is the ability to write assertions on relevant effects of the method under test.
If we cannot write assertions then we cannot check whether the behavior is correct or wrong.
Observability is affected by lack of accessibility in the sense of encapsulation, but also mockability (see Section~\ref{sec:mockability}).
\end{itemize}

\subsection{Quantitative Testability Measures} \label{sec:metrics}

Testability metrics have been well researched over the last
decades~\cite{Jun02,DBLP:journals/jss/BruntinkD06,DBLP:conf/icse/Zhao06,KM09,Cho09,DBLP:journals/sigsoft/KoutTB11,DBLP:journals/infsof/GarousiFK19,DBLP:conf/iwpc/TerragniSP20}.
The goal of these metrics is to find correlations between software
quality and design metrics and the difficulty to write tests.
Quantitative predictions can then be made about the effort required to
write tests.

Our goal, however, is to give precise diagnostic information, i.e.\ to
explain for each method where and what the concrete testability
problem is, assisting in fixing it and potentially aiming at fixing it
automatically. That way it is not only possible for a human to write
tests for their code, but also test generation tools will perform
better.

\begin{figure}[t]
\centering
\begin{tabular}{l@{\qquad}l}
\begin{lstlisting}
public class App {
  private static final logger = ...;
  private Client client;
  public App() {
    this.client = new Client();
  }
  public void send(Message m) {
    try {
      client.call(m);
    } catch (Exception e) {
      logger.error("send failed", e);
    }
  }
}
\end{lstlisting} &
\begin{lstlisting}
public class AppTest {
  @Test public void testSend() {
    // Arrange
    App app = new App();
    Message message = new Message("hello");
    // Act
    app.send(message);
    // Assert
    ???
  }
}
\end{lstlisting}
\end{tabular}
\caption{\label{fig:mockability-observability}
Observability aspect of mockability
}
\end{figure}

\begin{figure}[t]
\centering
\begin{tabular}{l@{\qquad}l}
\begin{lstlisting}
public class App {
  private static final logger = ...;
  private Client client;
  public App(Client client) {
    this.client = client;
  }
  public void send(Message m) {
    try {
      client.call(m);
    } catch (Exception e) {
      logger.error("send failed", e);
    }
  }
}
\end{lstlisting} &
\begin{lstlisting}
public class AppTest {
  @Test public void testSend() {
    // Arrange
    Client client = mock(Client.class);
    App app = new app(client);
    Message message = new Message("hello");
    // Act
    app.send(message);
    // Assert
    verify(client).call(message);
  }
}
\end{lstlisting}
\end{tabular}
\caption{\label{fig:mockability-observability-solved}
Observability aspect of mockability: observation through mock
}
\end{figure}

\section{Mockability} \label{sec:mockability}

We now make the notion of \emph{mockability} more explicit.
Mockability is the ability to inject objects that must be mocked in
order to control and observe their interactions.  Mockability is at
the core of the unit testability notion we use as it affects not only
controllability, but also observability.

In this section we will give examples to illustrate the
\emph{mockability} notion. Section~\ref{sec:model} will
then explain the underlying model.

To illustrate the controllability aspect,
Figure~\ref{fig:mockability-controllability} shows on the left
\textsf{Product} class and a method to add an expiry date.  The expiry
date is 30 days from now, and there is a method to check whether it
has expired.
A corresponding test would look as the one on the right.

The problem arises now when we want to test the is-expired-true case.
Obviously, we cannot just wait 30 days.
In order to test this properly, we need to mock the timestamps. 

This is shown in Figure~\ref{fig:mockability-controllability-solved}.
Fortunately, the Java library provides us with the \textsf{Clock}
class that allows us to control what \textsf{LocalDateTime.now}
returns. That means we can mock it, provided that we make the
dependency on \textsf{Clock} explicit so that we are able to inject
the mock.  What we can do then is to actually influence the behavior
through the mocked \textsf{Clock} and write a unit test for any test
case of the \textsf{isExpired} method without relying on the passage
of time.

Another quite common pattern is shown in
Figure~\ref{fig:mockability-observability}. We will use this example to
explain how mockability relates to observability.
We have an \textsf{App} class that has a client to send a message to
some service.  There are two problems here. First of all we do not want
to use the real service client; and second there is not really anything
to observe.

Figure~\ref{fig:mockability-observability-solved} shows that we first
have to make the client injectable and then mock the client in order
to control its behavior.
As a side effect of mocking the client we also have something to
assert, namely at least that the client was called with the message.
This would have been impossible with the real client.

In addition we can now also control the client to trigger the
exception path.  This is shown in
Figure~\ref{fig:mockability-observability-exception}.  However, we
cannot observe the exception because it has been swallowed.
In order to properly observe the exception we would have to do further
modifications to improve the observability.

These examples show very common problems in real world code that has
not been written with testability in mind from the very start.

\begin{figure}[t]
\centering
\begin{tabular}{l@{\qquad}l}
\begin{lstlisting}
public class App {
  private static final logger = ...;
  private Client client;
  public App(Client client) {
    this.client = client;
  }
  public void send(Message m) {
    try {
      client.call(m);
    } catch (Exception e) {
      logger.error("send failed", e);
    }
  }
}
\end{lstlisting} &
\begin{lstlisting}
public class AppTest {
  @Test public void testSendFailed() {
    // Arrange
    Client client = mock(Client.class);
    when(client.send(any()))
      .thenThrow(new Exception());
    App app = new App(client);
    Message message = new Message("hello");
    // Act
    app.send(message);
    // Assert
    assertThrows(Exception.class,
      () -> app.send(message));
    verify(client).call(message);
  }
}
\end{lstlisting}
\end{tabular}
\caption{\label{fig:mockability-observability-exception}
  Observability aspect of mockability: controlling exception cases,
  but lack of observability
}
\end{figure}

\section{Unit Testability Model} \label{sec:model}

\begin{figure}[t]
\centering  
\includegraphics[width=0.8\textwidth]{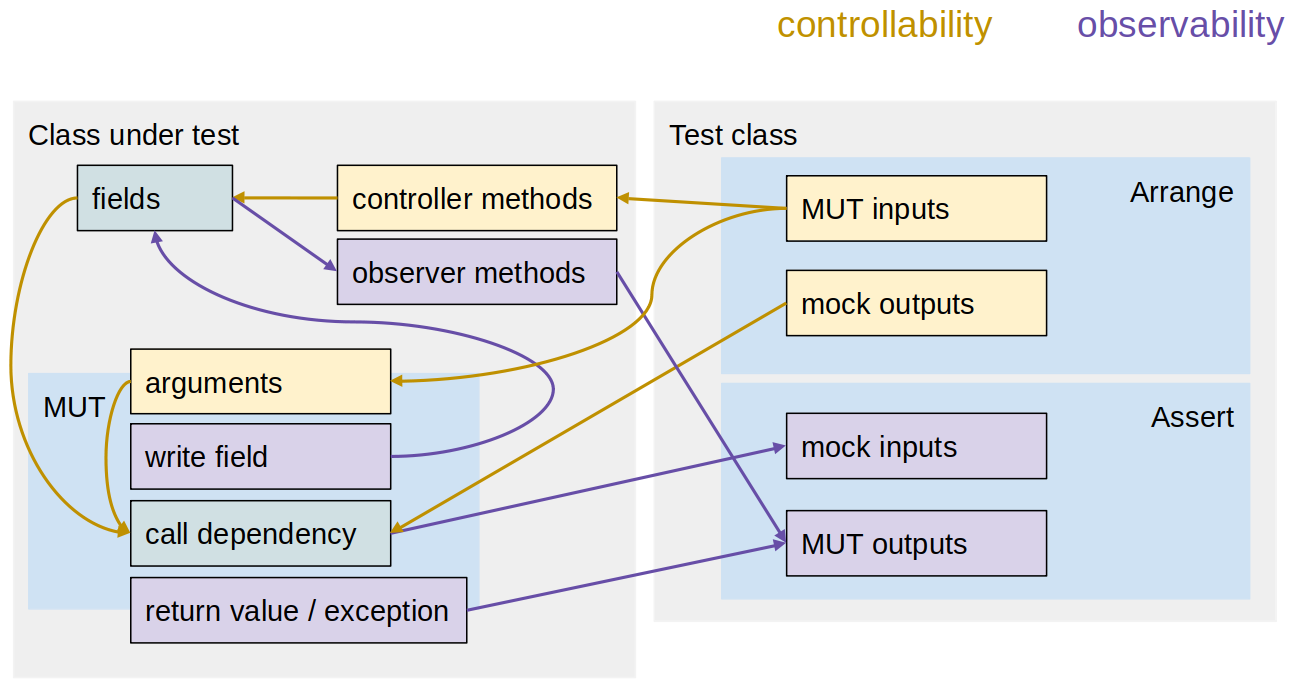}  
\caption{\label{fig:testability-model}
Unit Testability Model
}
\end{figure}

The examples given in the previous section show typical
unit testability issues occurring in real world code bases.
In order to understand formally what is behind these examples,
we need a model that describes the data flows involved.
This model will also be the basis for our unit testability
analysis outlined in the next section.

Figure~\ref{fig:testability-model} shows  on the left
the class under test and on the right the test class.

The arrows depict the data flows between the arrange and assert
sections and the method under test.
In yellow colors we have the controllability flows and in purple
the observability flows.
Fields and dependent components that may be called (in turquoise)
capture side effects.

\begin{itemize}
\item First the test arranges the inputs to the method under test. This is
to control the MUT by setting the fields through controller methods
such as setters and passing arguments to the MUT.  Also, we must be
able to inject mocks for dependencies that we may call.

\item Then the MUT may modify some fields.  If there are appropriate
observer methods such as getters then we can observe these side
effects and check them in the assert section.

\item When we call a mocked dependency then the arguments passed to the
mocked methods can be verified in the assert section.

\item The behavior of the mocked method is controlled in the arrange section.

\item Finally, the return value and exceptions, if any, can be observed and
checked in the assertion section.
\end{itemize}

If the controllability flow is broken somewhere then we have a
testability problem in the sense that we have trouble setting up an
appropriate arrange section and inject mocks.

If the observability flow is broken then we struggle to
write appropriate assertions that check the effects of
the execution of the MUT.

\section{Unit Testability Analysis} \label{sec:analysis}

We implemented a static analysis that analyzes the code on Java byte
code level.
The static analysis \emph{under-approximates} the set of methods that
are either not valuable to unit test or not unit-testable because of
controllability and observability problems.

The data flows that need to be considered in these static analyses
follow the model presented in Section~\ref{sec:model},
i.e. Figure~\ref{fig:testability-model}.

Several analyses are required that cooperate.
There is a mockability analysis that under-approximates the set of
non-mockable methods.
A method is called non-mockable if it
transitively calls methods that need to be mocked, but cannot.

Inductively this is defined as: A method is \emph{non-mockable} if either
\begin{itemize}
\item it must be mocked because it is non-deterministic, or
\item it has a call to a non-mockable static method, or
\item it calls a non-mockable instance method on an object that is non-injectable.
\end{itemize}
Then we need a second analysis
that under-approximates the \emph{non-injectable} objects, which are objects
that cannot be supplied through inputs.
For observability, we use analyses that follow similar definitions.

Note that by the nature of the under-approximate analyses that we use
we over-approximate the set of unit-testable methods, which may flag
fewer issues than there actually are.
The issues that we report are real instances of controllability
and observability issues, not potential ones.

\begin{figure}[t]
\centering  
\includegraphics[width=0.9\textwidth]{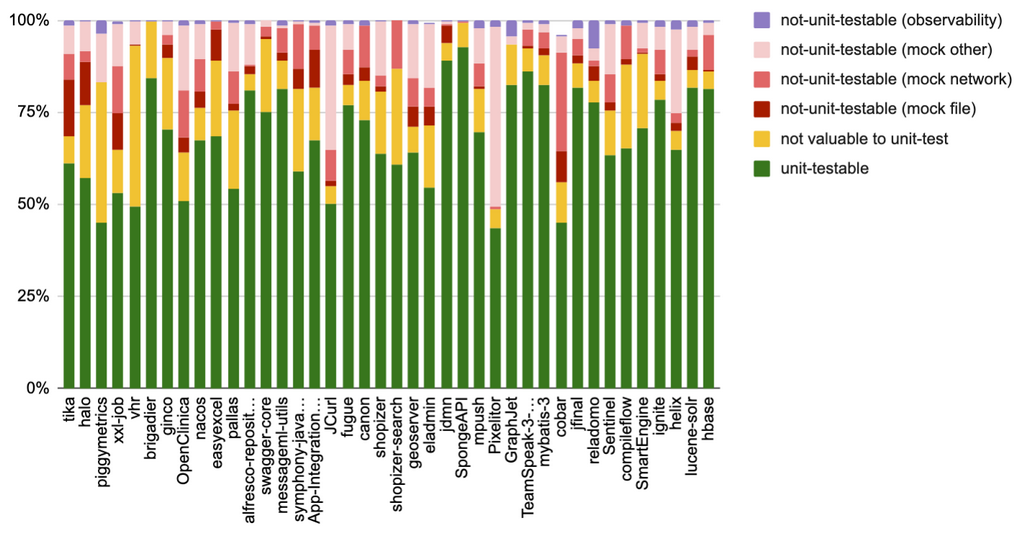}  
\caption{\label{fig:results-repositories}
Testability segmentation per repository
}
\end{figure}

\section{Experimental Results} \label{sec:results}

We analyzed 40 repositories with around 8 million lines of code.  The
dependencies that also needed to be analyzed increase the amount of code
analyzed by a further order of magnitude.  The repositories were open source
projects from various areas such as business workflows, data
processing, distributed computing and data storage.%
\footnote{The raw data is available at \url{https://bit.ly/2ZUNMOY}}

\subsection{Testability Maps}

Figure~\ref{fig:results-repositories} shows
the percentage of lines of code of each analyzed repository
according to testability classifications.

On average, 21\% of lines of code are not unit-testable.  This is for
various reasons that have been categorized in observability, mocking
of files, network and other system functionality, such as random or
time.
6\% are too trivial to be valuable to test,
and 73\% are unit-testable.

As we see there is quite some variability between projects.
Every project has its unique characteristics to some degree.
Small projects tend to be more extreme than larger ones, which are
closer to the mean.

On module level the variability is even larger.
Essentially, the various categories range from 0 to 100\%.  A
conjecture could be that code in the same module shares some
characteristics that make it exhibit similar testability aspects.

\begin{figure}[t]
\centering  
\includegraphics[width=0.9\textwidth]{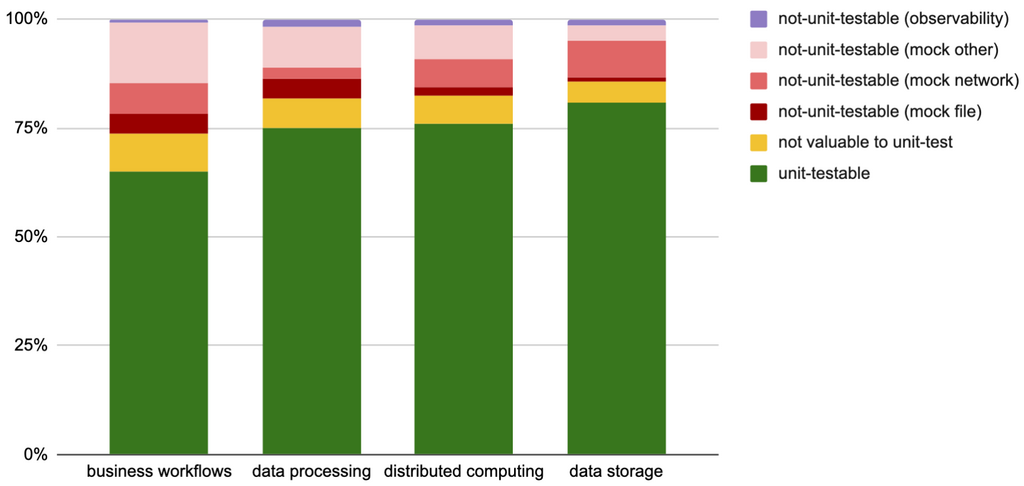}  
\caption{\label{fig:results-kinds}
Testability segmentation per application types
}
\end{figure}

Figure~\ref{fig:results-kinds} aggregates over different types of
applications.  There is not that much variation.  One could say that
data processing has proportionally more problems mocking file system
operations and data storage has more issues with networking.  Business
workflows seem slightly less testable.
One could speculate that these applications are
less mature than data storage systems.

We are less interested in statistics and their potential interpretations
here. The primary value of our analysis comes from the in-depth
\emph{diagnostic information} on the method level that can be provided.
From the obtained information we can build \emph{testability maps},
such as initially shown in Figure~\ref{fig:tika-testability-map}.
We will take a deeper look in the next too sections.

\subsection{Testability Assessment and Improvement} \label{sec:testability-assessment}

\begin{figure}[t]
\centering
\begin{tabular}{l@{\qquad}r}
\includegraphics[width=0.15\textwidth]{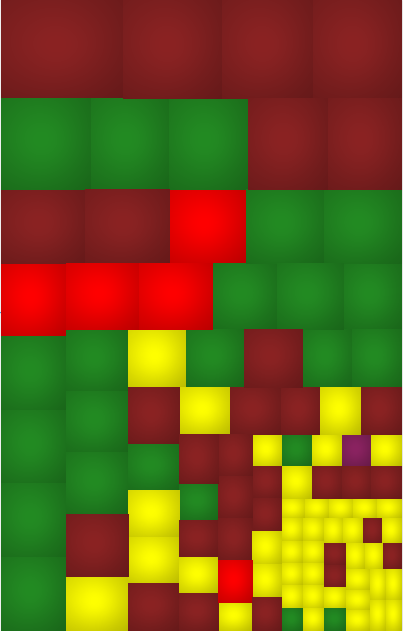} &
\includegraphics[width=0.15\textwidth]{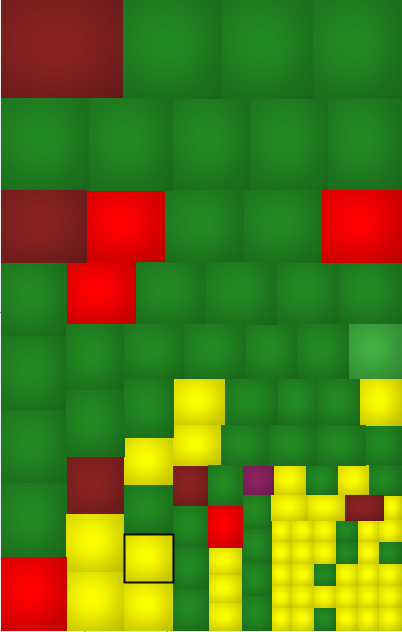}
\end{tabular}
\caption{\label{fig:testability-improvement}
Testability improvement: Before refactoring (left) and after (right)
}
\end{figure}

An application of testability maps is to spot potential areas of
concern in a code base and to perform targeted improvement of the
detected issues.

Figure~\ref{fig:testability-improvement} shows the testability of methods
in a code base before and after a refactoring.
The refactoring was performed before we had developed the testability
analysis. It was done due to bad performance of our test generation
tool.  The refactorings included improvements to injectability in
order to control non-determinism and improving observability of
relevant effects.
Figure~\ref{fig:testability-improvement} shows that these
refactorings, which significantly increased test generation
performance, are confirmed to improve testability by the testability
analysis in retrospect.
The number of unit-testable lines of code has increased from 36\% to 66\%.
\footnote{The methods in Figure~\ref{fig:testability-improvement} are
  ordered by LOCs, which have slightly changed during the refactoring,
  and thus do not exactly correspond in the left and right
  picture. Some methods have been reordered in this
  representation---unfortunate for the purpose of comparison.}

The availability of such a testability analysis could have allowed us
to explain to the customer immediately where and what the issues are
and they might even have been able to fix it themselves without our support.

\subsection{Diagnostic Testability Information}

\begin{figure}[t]

\centering
\includegraphics[width=0.7\textwidth]{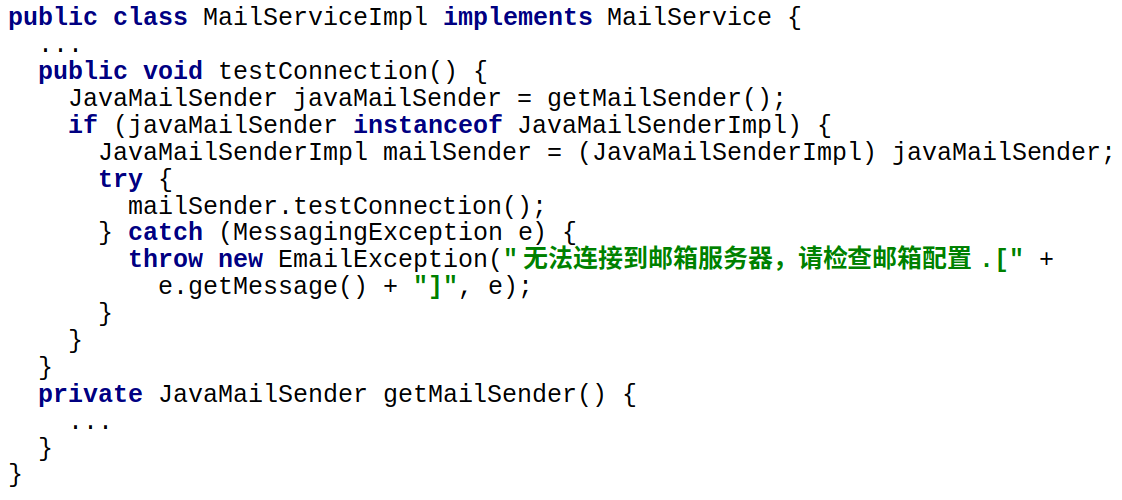}\\  
\includegraphics[width=0.7\textwidth]{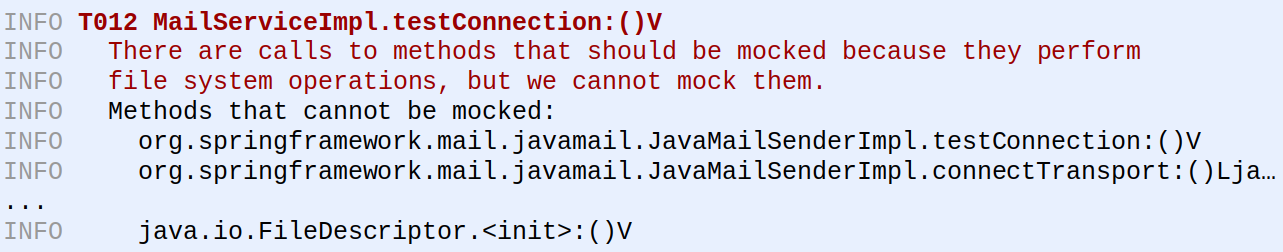}  
\caption{\label{fig:diagnostic-information}
  A method (taken from \url{https://github.com/halo-dev/halo})
  and corresponding diagnostic testability information
}
\end{figure}

In order to provide the developer advice how to resolve testability
issues, diagnostic information must be given. Such information is also
the first step in potentially automating testability refactorings.
Figure~\ref{fig:diagnostic-information} shows an example of
the kind of diagnostic information that we can produce at the moment.
The code example checks the connection to a mail server, which looks
quite similar to the client-server example in
Figure~\ref{fig:mockability-observability}.

When running the analysis on that class we get information that
the \textsf{testConnection} method calls methods that perform file system operations that cannot be mocked.
The diagnostic information shows a stack trace that leads to the root cause.
Information that is available, but not yet output is that the
\textsf{javaMailSender} instance is not injectable.

A possible resolution would be to make the \textsf{getMailSender}
method accessible in order to enable injection of a mock instance.
Making such a recommendation requires further analyses.

\section{Assumptions and Limitations} \label{sec:limitations}

The analysis described in Section~\ref{sec:analysis}
relies on lots of assumptions.
What is considered mockable and observable depends
on what is considered acceptable practice in unit tests.
For example, Java allows many dirty tricks to hack around restrictions.
For instance, reflection can be used to drill holes into encapsulated
objects. That way private methods can be called directly.
Whether this is acceptable depends on coding styles and habits,
how much maintainability is of a concern and probably also how big
the pressure is to complete work in time.
Another way are bytecode manipulations. This essentially means
rewriting binaries to make the code testable. Whether it is acceptable
that the code tested is not the code shipped depends on what the
alternatives are. It may not be considered recommended practice by
everyone.

Drawing the border between unit tests and integration tests is also
not always easy.
Talking to a database is probably not considered a unit test anymore.
Having multiple threads might be admissible sometimes.
Reading files may be fine, writing files is probably more debatable.
This is directly related to the question of what should be mocked.
Files system manipulations are not always that clear cut, 
network should probably be mocked, though.
Threaded code may have all sorts of issues.
Time and random generators usually require mocking.
But also in the latter case, there might be situations where it is
fine to have non-determinism as long as the test oracle remains
deterministic.

The guiding principle is always “what makes sense for
a developer”. However, as said, this may depend on specific
circumstances. Reasonable defaults and configurability
are hence required for such analyses in practice.

\section{Implications} \label{sec:implications}

This section describes some implications and learnings that
can be drawn from the presented study.

\subsection{What are the implications on testing efficiency?}

\begin{figure}[t]
\centering  
\includegraphics[width=0.25\textwidth]{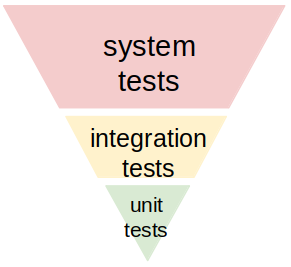}  
\caption{\label{fig:test-pyramid-upside-down}
Unbalanced testing pyramid due to lack of unit testability
}
\end{figure}

Lack of unit-testability leads to an imbalanced testing pyramid,
see Figure~\ref{fig:test-pyramid-upside-down} in comparison with
the desired test composition depicted in Figure~\ref{fig:test-pyramid}.
There tends to be a higher proportion of system and integration tests.
It is very tempting to write such tests when code is entangled and
thus unit tests require a lot of effort and intricate setup.
However, this leads to slower CI runs with all its consequences.
Developers start complaining and consider only running some of the
tests on each pull request and the remaining ones nightly.  When those
nightly tests fail then someone has to spend extra effort in fixing
those bugs that would otherwise never have made it into the code base.
So, the expected shift-left of defect detection is not possible with
such an approach.

W.r.t. code coverage, it is important to focus on critical code
coverage.  In particular in projects that have a high percentage of
trivial code.  Tricking coverage figures by 10\% and more is an easy
exercise otherwise.
Many companies have code coverage goals, such as 80\%, which are
pretty meaningless if the remaining 20\% are critical code that really
needs to be tested.
Therefore it is essential to focus on the critical code.  The
testability analysis discussed in Section~\ref{sec:analysis} can be
used for normalization and also for pointing to areas of risk in the code
base that need special attention.

\subsection{What is the role of design for testability?}

As mentioned in Section~\ref{sec:limitations},
there are common workarounds to make Java code more unit-testable.
Bytecode rewriting is one of them. Another way out is not to write
unit-tests, but integration tests, by testing against a real database,
for instance.
Of course, this slows down the build and fails to deliver on the goal
to shift left defect detection in the development cycle.
Reflection is a possible workaround to observability issues. Sometimes
also ad-hoc weakenings of access modifies are performed.
It is not uncommon that assertions are written on the content
of log files.

All this does not really improve the software quality and leads to
further problems down the road.  What actually should be done is to
consider testing early in the development process and to plan for the
testing requirements when designing the interfaces.
Nobody would design a piece of hardware without building in facilities
for testing it.  The integration of testing and debugging support into
a hardware system, such as JTAG~\cite{jtag13}, needs to be planned
from the beginning. We do not routinely think in terms of testing
interfaces about software yet, unfortunately.

When talking to companies, test-first approaches such as test-driven
development are quite popular and many teams want to adopt them.  Of
course, even on successful adoption, this has only an immediate impact
on fresh code whereas the effects will show much more slowly on the big
pile of existing software.

Also moving to application frameworks that support dependency
injection such as Spring~\cite{spring20} or Guice~\cite{guice20} could
generally be expected to improve testability.

\subsection{What are the implications on automated verification tools?}

\begin{wrapfigure}{r}{0.2\textwidth}
\includegraphics[width=0.15\textwidth]{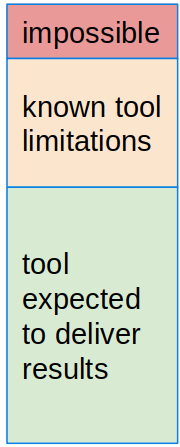}  
\caption{\label{fig:verifiability-map}
Verifiability map of a code base
}
\end{wrapfigure}

When automated verification tools such as model checkers are applied
to large software then this is usually done in a divide-and-conquer
fashion.
The verification harnesses that are needed to verify the individual
parts are very similar to unit tests, except that they do not use
concrete values in the arrange section, but non-determinism and
assumptions; and the assertions are usually more general
postconditions.
Automated test generation tools essentially produce such harnesses
automatically and instantiates them concretely.
Unless a verification tool has an almighty view on the unit under test
the problems are similar.
Code designed for testability can be expected to be easier to handle.

Similarly to the testability analysis discussed in
Section~\ref{sec:analysis} one could have something like a
\emph{verifiability analysis} which, for instance, maps out a code
base into parts that are practically impossible to handle, parts that
we know the tool cannot handle because there are known limitations,
and parts that the tool is expected to fairly reliably deliver results
on (see Figure~\ref{fig:verifiability-map}).

In the case of automated test generation, testability maps
(cf. Section~\ref{sec:testability-assessment}) are a tool for allowing
the user to quickly spot upfront where issue are likely to be expected
and thus manage expectations early.
For a verification tool, when applied to large projects, a
verifiability analysis could give some indication on where to start
and what the upper bounds are on what the verification tool can
be expected to achieve.

\section{Conclusions and Prospects} \label{sec:conclusions}

Business software is business-critical and there is a drive to
increase unit-testing levels in order to be faster than the
competition.
Engineering managers and software architects primarily care about
testing the critical parts of the code and thus test coverage needs to
be normalized to those.
Not all code is equally easy to test and a testability analysis can
determine where issues with controllability and observability are
present in a code base and why.
Testability issues impact automated verification tools such as model
checkers and test generators.

Diagnosing unit testability issues precisely is the first step.  The next
step is to provide automated resolution to improve testability by
automated refactorings.

\bibliographystyle{splncs04}
\bibliography{bibliography}

\end{document}